\documentclass[epj]{svjour}
\usepackage{latexsym}
\usepackage{graphicx}
\usepackage{amsmath}
\usepackage{hyperref}

\begin{document}

\title{Exact dynamics of two ultra-cold bosons confined in a one-dimensional double-well potential}
\subtitle{}
\author{Jacek Dobrzyniecki\inst{1} \and Tomasz Sowi\'nski\inst{2}
} 

\institute{
Faculty of Mathematics and Natural Sciences, Cardinal Stefan Wyszy\'nski University \\
ul. W\'oycickiego 1/3, 01-938 Warsaw, Poland \and 
Institute of Physics of the Polish Academy of Sciences \\ 
Al. Lotnik\'ow 32/46, 02-668 Warsaw, Poland 
}

\date{Accepted: March 8, 2016}
\mail{tomsow@ifpan.edu.pl}

\abstract{
The dynamics of two ultra-cold bosons confined in a one-dimensional double-well potential is studied. 
We compare the exact dynamics governed by a full two-body Hamiltonian with the dynamics obtained in a two-mode model approximation. We show that for sufficiently large interactions the two-mode model breaks down and higher single-particle states have to be taken into account to describe the dynamical properties of the system correctly. 
} 

\titlerunning{Exact dynamics of two bosons in a double-well potential} \authorrunning{J. Dobrzyniecki and T. Sowi\'nski}

\maketitle

\section{Introduction}
Double-well confinement is one of the simplest examples where quantum dynamics manifests its nonintuitive properties \cite{GriffBook,MerzbBook}. Already on the single-particle level, the tunneling through a classically impenetrable barrier has many spectacular consequences like electron tunneling through p-n junctions \cite{Lashkaryov,Ohl,Shockley,Esaki} or the Josephson effect \cite{Josephson}. The physics of double-well systems becomes even more interesting whenever interactions between particles are considered.

In view of recent experimental progress with ultra-cold atoms forming a Bose-Einstein condensate, double-well systems are one of the most commonly exploited schemes studied \cite{Andrews,Smerzi,Milburn,Menotti,Meier,Shin,Albiez,Levy,Salgueiro,Simon,Liu}. Typically, in this context one assumes that weakly interacting bosons occupying different wells can be described with two independent single-particle orbitals and that the dynamics is governed by two mechanisms: contact two-body interactions acting locally and single-particle tunneling between wells. Then, in the mean-field limit, a corresponding Gross-Pitaevskii equation is introduced and numerically solved for different initial conditions \cite{Raghavan,Ostrovskaya,Ananikin}. Generalized two-mode models, taking into account additional terms originating from long-range interactions or occupation-dependent tunnelings, are also considered in the literature and relevant corrections to the dynamics are studied \cite{Lahaye,Adhikari,Bruno}. Although the validity of these simplified two-mode models was confirmed experimentally for weak interactions between particles, they were extended beyond the range of their applicability and adopted for strongly interacting systems, i.e. in situations when the local interaction energy is much larger than the single-particle tunneling energy. For example, it was shown that for initially imbalanced occupations the dynamics is heavily affected by strong interactions \cite{DuttaS}. Unfortunately, the validity of the model used was not discussed and its predictions were not compared with the exact dynamics governed by a general model. 

It is quite obvious that for sufficiently strong interactions between particles any two-mode model has to break down. This comes from the observation that interactions always introduce some multi-particle correlations that exist locally at each site of the potential. Therefore, a simple assumption that all particles occupying a single site can be described with a single effective orbital cannot be valid. The situation is very similar to the problem of ultra-cold bosons confined in a single harmonic trap. As it was shown in the case of two strongly interacting bosons in a harmonic trap, a single-mode description is not valid \cite{Sowinski2010}. In the case studied here, whenever interactions are significantly larger than typical tunneling energies between wells, local correlations induced by interactions are produced much faster than correlations between wells. Just from this simple observation it is quite obvious that any two-mode approximation is inconsistent.

Inspired by this simple observation, in this article we study the dynamical properties of two bosons confined in a one-dimensional double-well potential and initially occupying a chosen site. We numerically compare the exact, many-body dynamics of the system with the dynamics governed by simplified two-mode Hamiltonians. The comparison is performed for different interaction strengths and different depths of the modeled double-well potential.

The paper is organized as follows. In Section 2 the single-particle Hamiltonian with a double-well potential is introduced and its spectral properties are described. In Section 3 we discuss the full many-body Hamiltonian describing identical and spinless ultra-cold bosons and we rewrite it in a two-site multi-orbital Bose-Hubbard form. Then, in Section 4 we focus on the problem of two bosons and we explain the simplifications of the multi-orbital Hamiltonian commonly used in literature. In Section 5 the main results are presented. We compare the evolution governed by a general Hamiltonian with those predicted by simplified models in different regimes of interactions and double-well depths. Finally, we conclude in Section 6.

\section{Double-well model}
\begin{figure}
\includegraphics{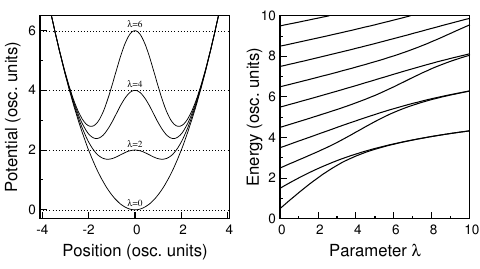}
\caption{Left: The shape of the potential $V(x)$ for different values of the parameter $\lambda$. For $\lambda=0$ the standard harmonic oscillator shape is restored. For larger $\lambda$, two symmetric minima are separated by a potential barrier. Right: The spectrum of the single-particle Hamiltonian \eqref{Schrodinger} as a function of the double-well parameter $\lambda$. For $\lambda=0$, the standard spectrum of the harmonic oscillator is obtained. For larger $\lambda$'s two-fold degeneracy between even and odd eigenstates appear. In both figures, all the quantities are given in harmonic oscillator units, i.e., energy in $\hbar \Omega$ and position in $\sqrt{\hbar/m\Omega}$.\label{Fig1}}
\end{figure}

For clarity, in this article we assume that particles of mass $m$ are confined in a one-dimensional harmonic potential with frequency $\Omega$. The double-well configuration is forced by an additional repulsive gaussian potential with a controlled intensity which is centered in the middle of the trap. The external potential is modeled by the following function:
\begin{equation} 
V(x) = \hbar\Omega\left[\frac{m\Omega}{2\hbar}x^2 + \lambda\, \mathrm{exp}\left(-\frac{m\Omega}{2\hbar}x^2\right)\right].
\end{equation}
The dimensionless parameter $\lambda>0$ gives the intensity of the barrier between double-well sites (See the left panel in Fig. \ref{Fig1}). Depending on $\lambda$, the complete spectrum of the corresponding single-particle Schr\"odinger equation
\begin{equation} \label{Schrodinger}
H_0\varphi_i(x) = {\cal E}_i \varphi_i(x),
\end{equation}
where $H_0=-\frac{\hbar^2}{2m}\frac{\mathrm{d}^2}{\mathrm{d}x^2} + V(x)$, is found numerically via an exact diagonalization procedure. The diagonalization is done in the position representation on a dense grid. The resulting spectrum as a function of $\lambda$ is presented in the right panel of Fig. \ref{Fig1}. Obviously, for $\lambda=0$ the standard spectrum of the harmonic oscillator is obtained. For increasing $\lambda$ the single-particle spectrum changes and a characteristic two-fold quasi-degeneracy between even and odd eigenstates appears. For double-well problems it is convenient to introduce another single-particle basis $\{\phi_{Li}(x),\phi_{Ri}(x)\}$ of states localized in a given well. The basis is constructed directly form the eigenstates of the Hamiltonian \eqref{Schrodinger} as follows:
\begin{subequations}\label{LeftRightBasis}
\begin{align} 
\phi_{Li}(x) &= \frac{\varphi_{2i}(x)-\varphi_{2i+1}(x)}{\sqrt{2}}, \\
\phi_{Ri}(x) &= \frac{\varphi_{2i}(x)+\varphi_{2i+1}(x)}{\sqrt{2}}.
\end{align}
\end{subequations}
Obviously, the states $\{\phi_{\sigma i}(x)\}$, where the index $\sigma=\{L,R\}$ enumerates wells of the potential, are not eigenstates of the Schr\"odinger equation \eqref{Schrodinger}. However, for each state $\phi_{\sigma i}(x)$ there is only one off-diagonal term coupling the state with the state localized in the other site. The corresponding matrix elements of the Hamiltonian $J_i = -\int\phi^*_{Li}(x)H_0\phi_{Ri}(x)$ are called tunnelings. Moreover, just from the construction it follows that the diagonal terms $E_i = \int\phi^*_{\sigma i}(x)H_0\phi_{\sigma i}(x)$ do not depend on the site index $\sigma$. The tunnelings $J_i$ together with energies $E_i$ are related to eigenvalues ${\cal E}_i$ in equation \eqref{Schrodinger}:
\begin{equation} \label{tuneling}
J_{i} = \frac{{\cal E}_{2i+1}-{\cal E}_{2i}}{2}, \qquad E_{i}=\frac{{\cal E}_{2i+1}+{\cal E}_{2i}}{2}.
\end{equation}

\section{Many-body Hamiltonian}
In the following we consider two ultra-cold bosons confined in the potential $V(x)$ and interacting via short-range forces modeled with a point-like potential $g\delta(x-x')$, where $g$ is an effective interaction constant related to the $s$-wave scattering length. Note that, in contrast to higher dimensions, in the one-dimensional case the Dirac $\delta$ function is a well defined self-adjoint operator and any regularization is not necessary \cite{Bush}. Although we consider only two particles, it is still convenient to work in the formalism of the second quantization. Therefore, we introduce the field operator $\hat\Psi(x)$ annihilating a particle at position $x$. The field operator fulfills natural bosonic commutation relations $\left[\hat\Psi(x),\hat\Psi^\dagger(x')\right]=\delta(x-x')$ and $\left[\hat\Psi(x),\hat\Psi(x')\right]=0$. In this language the many-body Hamiltonian of the system can be written in the form 
\begin{equation}\label{Hamiltonian}
\hat{\cal H}=\int\!\!\mathrm{d}x\left[ \hat\Psi^\dagger(x)H_0\hat\Psi(x)+\frac{g}{2}\hat\Psi^\dagger(x)\hat\Psi^\dagger(x)\hat\Psi(x)\hat\Psi(x)\right].
\end{equation}
The standard route to analyze the Hamiltonian \eqref{Hamiltonian} is to decompose the field operator $\hat\Psi(x)$ in a chosen single-particle basis. Here we decompose the bosonic field in the basis of left-right single-particle states of the double-well potential
\begin{equation} \label{decompose}
\hat\Psi(x) = \sum_\sigma\sum_i \hat{a}_{\sigma i}\phi_{\sigma i}(x),
\end{equation}
where an operator $\hat{a}_{\sigma i}$ annihilates a boson at site $\sigma$ in level $i$, i.e. a boson in a single-particle state described by the wave function $\phi_{\sigma i}(x)$. These operators fulfill standard bosonic commutation relations $\left[\hat{a}_{\sigma i},\hat{a}^\dagger_{\sigma' j}\right]=\delta_{\sigma\sigma'}\delta_{ij}$ and $\left[\hat{a}_{\sigma i},\hat{a}_{\sigma' j}\right]=0$. After substitution of the decomposition \eqref{decompose} into \eqref{Hamiltonian} one rewrites the Hamiltonian in a Hubbard-like form
\begin{align} \label{HamHubbard}
\hat{\cal H} &= \sum_\sigma\sum_i E_i\hat{a}^\dagger_{\sigma i}\hat{a}_{\sigma i}-\sum_i J_i(\hat{a}^\dagger_{Li}\hat{a}_{Ri}+\hat{a}^\dagger_{Ri}\hat{a}_{Li})  \nonumber \\
&+\frac{1}{2}\sum_{ABCD}U_{ABCD}\hat{a}^\dagger_{A}\hat{a}^\dagger_{B}\hat{a}_{C}\hat{a}_{D},
\end{align}
where for simplicity we have introduced in the interaction part of the Hamiltonian a super-index $A=(\sigma,i)$ indicating two quantum numbers: site index $\sigma$ and excitation index $i$ respectively. Values of the interaction terms $U_{ABCD}$ can be calculated straightforwardly from the shapes of single-particle states
\begin{equation}
U_{ABCD} = g\int\mathrm{d}x\,\phi_{A}^*(x)\phi_{B}^*(x)\phi_{C}(x)\phi_{D}(x).
\end{equation}
The Hamiltonian \eqref{HamHubbard} is completely equivalent to the initial Hamiltonian \eqref{Hamiltonian}. To make the notation simpler, in what follows we also introduce operators of the total number of bosons in a given energy level $\hat{n}_i$ and the total number of bosons in a given site $\hat{N}_\sigma$:
\begin{subequations}
\begin{align}
\hat{n}_i&= \hat{a}^\dagger_{L i}\hat{a}_{L i}+\hat{a}^\dagger_{R i}\hat{a}_{R i}, \\
\hat{N}_L &= \int_{-\infty}^{0}\!\!\mathrm{d}x\,\Psi^\dagger(x)\Psi(x), \\
\hat{N}_R &= \int_{0}^{\infty}\!\!\mathrm{d}x\,\Psi^\dagger(x)\Psi(x).
\end{align}
\end{subequations}
Note that the operators $\hat{N}_\sigma$ are defined very differently to a common definition of a sum over occupations $\sum_i \hat{a}^\dagger_{\sigma i}\hat{a}_{\sigma i}$. In our definition, the spreading of the basis wave functions $\phi_{\sigma i}(x)$ to the neighboring well is automatically taken into account. Moreover, the definition used here treats single-particle superpositions appropriately, i.e. a left/right probability is calculated directly from the full single-particle density and not as a sum of partial probabilities.

\section{The system studied}
To study the dynamical properties of the system we assume that initially two bosons occupy the lowest state of a chosen (left) site of the double-well potential
\begin{equation} \label{Ini}
|\mathtt{ini}\rangle=\frac{1}{\sqrt{2}}\hat{a}_{L0}^{\dagger 2}|\mathtt{vac}\rangle.
\end{equation}
It is worth noticing that this kind of state can be prepared experimentally for a few particles. For example, a similar state for two distinguishable particles was obtained in recent experiments \cite{Jochim}. Therefore, the problem analyzed here has not only a theoretical meaning but also may have some importance in few-body problems considered recently in the field of ultra-cold atomic systems \cite{Blume,Zinner}. It is also worth noting that analogous systems were studied recently with engineered optical waveguide lattices \cite{Longhi,Longhi2}.

The initial state $|\mathtt{ini}\rangle$ is not an eigenstate of the many-body Hamiltonian \eqref{HamHubbard} and its time evolution is not trivial. Our aim is to compare the exact dynamics of the state governed by the full many-body Hamiltonian \eqref{HamHubbard} with the dynamics predicted by a simplified model in a two-mode approximation, i.e. when the decomposition \eqref{decompose} is cut down to the two lowest single-particle states, $i=0$. In this approximation the Hamiltonian \eqref{HamHubbard} has the form
\begin{align} \label{Ham2Mode}
\hat{\cal H}_{\mathtt{2Mode}} &= -J (\hat{a}^\dagger_{L0}\hat{a}_{R0}+\hat{a}^\dagger_{R0}\hat{a}_{L0}) \nonumber \\
&+
\frac{U}{2}\left(\hat{a}^{\dagger 2}_{L 0}\hat{a}_{L 0}^2+\hat{a}^{\dagger 2}_{R 0}\hat{a}_{R 0}^2\right)+V \hat{a}^\dagger_{L0}\hat{a}_{L0}\hat{a}^\dagger_{R0}\hat{a}_{R0}
\nonumber \\
&+T \left(\hat{a}^\dagger_{L0}\,\hat{n}_{0}\,\hat{a}_{R0}+\hat{a}^\dagger_{R0}\,\hat{n}_{0}\,\hat{a}_{L0}\right) \nonumber \\
&+\frac{V}{4}\left(\hat{a}^{\dagger 2}_{L0}\hat{a}_{R0}^2+\hat{a}^{\dagger 2}_{R0}\hat{a}_{L0}^2\right),
\end{align}
where for simplicity we introduce standard notations for relevant parameters:
\begin{subequations} \label{ParamInt}
\begin{align}
U &=g\int\!\!\mathrm{d}x\,\left[\phi_{L0}(x)\right]^4=g\int\!\!\mathrm{d}x\,\left[\phi_{R0}(x)\right]^4, \\
V &=2g\int\!\!\mathrm{d}x\,\left[\phi_{L0}(x)\phi_{R0}(x)\right]^2, \\
T &=g\int\!\!\mathrm{d}x\,\phi_{L0}(x)\left[\phi_{R0}(x)\right]^3,
\end{align}
\end{subequations} 
and $J=J_0$ defined in \eqref{tuneling}. Note that in definitions \eqref{ParamInt} we directly exploit the fact that single-particle wave functions are real functions of positions. The additional interaction-induced tunneling controlled by the amplitude $T$ depends directly on the overlap between wave functions localized in neighboring wells. Therefore, it becomes important for shallow barriers.

It is also worth noticing that in the model studied, pair-tunneling and inter-site density-density interaction terms are controlled by the same parameter $V$. This is a direct consequence of the short-range interactions that are assumed. Therefore, a typical approximation made in this context (see for example \cite{DuttaS}) of neglecting pair-tunneling without neglecting density-density interactions seems to be inconsistent. 

It is known that in the case of an extended Hubbard model describing bosons interacting via long-range interactions, additional tunnelings, i.e. pair-tunneling and density-induced tunneling, may play a crucial role and may lead to the appearance of exotic quantum many-body phases in the system \cite{Dutta2011,Sowinski2012,DuttaReview}. 

Typically, on this level of simplification, further approximations are performed and terms controlled by amplitudes $T$ and $V$ are neglected. Then, the Hamiltonian is reduced to the standard two-site Bose-Hubbard form \cite{Milburn,Diaz}
\begin{align} \label{HamRed}
\hat{\cal H}_{\mathtt{R}} &= -J (\hat{a}^\dagger_{L0}\hat{a}_{R0}+\hat{a}^\dagger_{R0}\hat{a}_{L0}) 
+
\frac{U}{2}\left(\hat{a}^{\dagger 2}_{L 0}\hat{a}_{L 0}^2+\hat{a}^{\dagger 2}_{R 0}\hat{a}_{R 0}^2\right).
\end{align}

In the case of two ultra-cold bosons the simplified models \eqref{Ham2Mode} and \eqref{HamRed} have a very simple matrix representation. Indeed, in this case the Hilbert space is spanned by three two-body Fock states $|20\rangle=\frac{1}{\sqrt{2}}\hat{a}_{L0}^{\dagger 2}|\mathtt{vac}\rangle$, $|02\rangle=\frac{1}{\sqrt{2}}\hat{a}_{R0}^{\dagger 2}|\mathtt{vac}\rangle$, and $|11\rangle=\hat{a}_{L0}^{\dagger}\hat{a}_{R0}^{\dagger}|\mathtt{vac}\rangle$. Therefore, the Hamiltonian \eqref{Ham2Mode} can be rewritten in a simple $3\times 3$ matrix form 
\begin{equation} \label{Ham2ModeMatrix}
\hat{\cal H}_{\mathtt{2Mode}} = \left(
\begin{array}{ccc}
U & \sqrt{2}(T-J) & V/2 \\
\sqrt{2}(T-J) & V & \sqrt{2}(T-J) \\
V/2 & \sqrt{2}(T-J) & U 
\end{array}\right).
\end{equation}
The matrix form of the reduced Hamiltonian \eqref{HamRed} can be derived straightforwardly from \eqref{Ham2ModeMatrix} by setting $T=V=0$.  

It is worth noticing that from a theoretical point of view, the latter model \eqref{HamRed} is controlled effectively by only one dimensionless parameter $U/J$. However, from experimental side the two parameters $U$ and $J$ are controlled independently, i.e. the single-particle tunneling $J$ is controlled only by an external potential parameter $\lambda$; the on-site interaction $U$ is controlled by $\lambda$ and a coupling strength $g$ which can be tuned independently. That means that in this approximation the same value of the dimensionless parameter $U/J$ can be obtained for different parameters $\lambda$ by changing the coupling $g$. Since $U/J$ is the only parameter in the model, in all these cases the evolution of the system is the same provided that time is measured in the units of $\hbar/J$. Of course for shallow barriers, when the interaction energy is comparable with the energy gap to higher orbitals, terms neglected in the original Hamiltonian start to influence the dynamics and an evolution for the same ratio $U/J$ starts to depend on $\lambda$. This would be the sign that the simplified model breaks down.

\section{The dynamics}
To analyze the limitations of the simplifications of the Hamiltonian, we study the dynamics governed by Hamiltonians \eqref{HamHubbard}, \eqref{Ham2Mode}, and \eqref{HamRed} of the system initially prepared in the state $|\mathtt{ini}\rangle$. In our numerical approach we first diagonalize the matrix forms of Hamiltonians in the Fock basis built from single-particle states \eqref{LeftRightBasis} and we obtain two-body eigenstates and their eigenenergies. In the case of the full multi-orbital Hamiltonian \eqref{HamHubbard} we cut the decomposition \eqref{decompose} at a sufficiently large $i_{\mathrm{max}}$ (in practice $i_{\mathrm{max}}=30$). Then we decompose the initial state $|\mathtt{ini}\rangle$ into the eigenstates and we find the time-dependent state of the system $|\psi(t)\rangle$. In principle, an exact form of the state $|\psi(t)\rangle$ depends on the approximation assumed.

The simplest quantity that is accessible in experiments and can be compared for different theoretical models is an expectation value of the site occupation number $\langle \psi(t)|\hat{N}_\sigma|\psi(t)\rangle$. Since the total number of particles is conserved, in practice it is sufficient to calculate the imbalance of populations between potential wells
\begin{equation}
I(t) = \langle \psi(t)|\hat{N}_L - \hat{N}_R|\psi(t)\rangle.
\end{equation} 

In the following, we compare the dynamics of the system in two different regimes of parameters. To make the comparison clear we compare two different barriers: $\lambda=3$ and $\lambda=5$. In both cases we tune the interaction coupling $g$ in such a way that the ratio $U/J$ represents three different regimes: soft interactions ($U/J=0.1$), intermediate interactions ($U/J=1$), and strong interactions ($U/J=12$). With this strategy, a comparison of different cases is quite natural since simplified Hamiltonian \eqref{HamRed} depends only on the ratio $U/J$ (with $J$ fixing the energy scale) and therefore it gives exactly the same dynamics for both $\lambda$'s (understood as an evolution of coefficients in the decomposition of the many-body state in a Fock state basis). Whereas, in the other cases, there is also an explicit dependence on $\lambda$ through the other parameters of the Hamiltonian, $T$ and $V$.

\begin{figure} 
\includegraphics{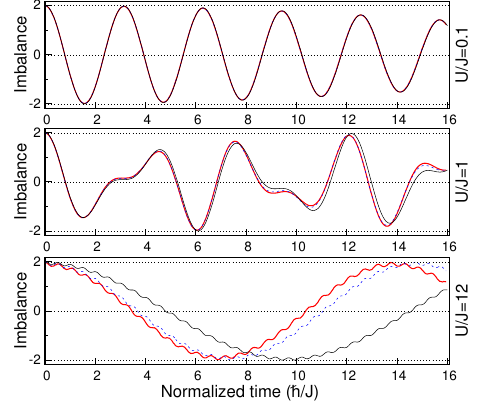}
\caption{Imbalance of particles $I(t)=n_L(t)-n_R(t)$ as a function of normalized time in the case of a deep potential barrier ($\lambda=5$). The solid red curve corresponds to the dynamics governed by a full multi-orbital Hamiltonian \eqref{HamHubbard}. The dotted blue and thin black curves show the simplified dynamics described by Hamiltonians \eqref{Ham2Mode} and \eqref{HamRed}, respectively. In all cases we assume that initially the system is prepared in the state $|\mathtt{ini}\rangle$. Note that the exact dynamics is well approximated by the reduced Hamiltonians independently of the strength of the interactions being considered.\label{Fig2}}
\end{figure}

In Fig. \ref{Fig2} we show the evolution of imbalance $I(t)$ in the case of a very deep potential barrier, $\lambda=5$. The solid red line represents the imbalance of particles predicted by an exact many-body Hamiltonian \eqref{HamHubbard}. The dotted blue and thin black lines correspond to the simplified two-mode models \eqref{Ham2Mode} and \eqref{HamRed}, respectively. In this case, independently of the strength of interactions between particles, both simplified models recover the exact dynamics appropriately. A small phase shift in oscillations is visible only in the very strong interaction regime. However, the general behavior of the imbalance is reproduced. 

\begin{figure} 
\includegraphics{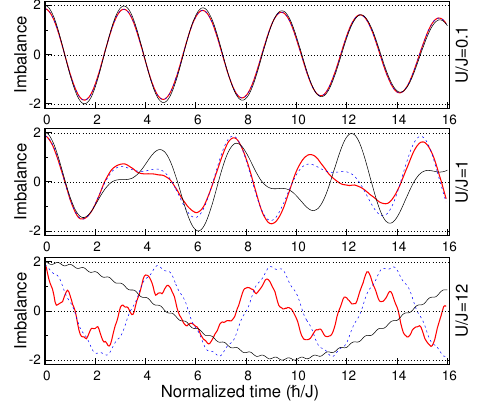}
\caption{Imbalance of particles $I(t)=n_L(t)-n_R(t)$ as a function of normalized time in the case of a shallow potential barrier ($\lambda=3$). The solid red curve corresponds to the dynamics governed by a full multi-orbital Hamiltonian \eqref{HamHubbard}. The dotted blue and thin black curves show the simplified dynamics described by Hamiltonians \eqref{Ham2Mode} and \eqref{HamRed}, respectively. In all cases we assume that initially the system is prepared in the state $|\mathtt{ini}\rangle$. In contrast to the cases shown in Fig. \ref{Fig2}, the exact dynamics can not be approximated by the most simplified Hamiltonian \eqref{HamRed}. However, comparison of the solid red and dotted blue lines may suggest that the dynamics of occupations are well captured by the two-mode Hamiltonian \eqref{Ham2Mode}. As explained in the main text and in Fig. \ref{Fig4}, the two-mode Hamiltonian is not sufficient to describe correlations between particles. \label{Fig3}}
\end{figure} 
The situation changes significantly for a shallow barrier. In Fig. \ref{Fig3} we compare the analogous behavior for $\lambda=3$. It is seen that only in the case of small interactions is the dynamics well captured by simplified models. However, a comparison of the dotted blue and solid red lines suggests that a full two-mode model \eqref{Ham2Mode} works quite well even for strong interactions and that it can predict the evolution of the imbalance of occupations in a satisfactory way for long times. Only the small oscillations around the general behavior are not reproduced correctly by this model. For example, for $U/J=12$, where the simplified model \eqref{HamRed} is completely wrong, the model \eqref{Ham2Mode} is sufficient to describe the oscillation of occupation between different wells with almost adequate frequency. The small discrepancy in the resulting frequency leads only to a slow growth of the phase shift between curves. 
 
This illusory conviction that a complete two-mode Hamiltonian \eqref{Ham2Mode} is sufficient to describe the dynamical properties of the system in the strong interaction limit has to be revisited when, instead of densities, inter-particle correlations are considered. For example, let us consider one of the simplest correlations -- the probability that bosons occupy different wells of the potential. In the case of two bosons, this probability is related to the density-density correlation: 
\begin{equation} \label{Prob}
{\cal P}(t) = \sum_{ij}\langle \psi(t)|\hat{n}_{L j}\hat{n}_{R i}|\psi(t)\rangle,
\end{equation}
where $\hat{n}_{\sigma i}=\hat{a}^{\dagger}_{\sigma i}\hat{a}_{\sigma i}$.

\begin{figure} 
\includegraphics{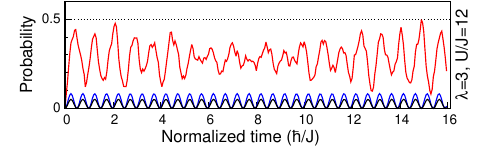}
\caption{Time evolution of the probability that bosons occupy different wells of the potential in the case of a shallow potential barrier ($\lambda=3$) and strong interactions ($U/J=12$). For simplified Hamiltonians \eqref{Ham2Mode} and \eqref{HamRed} (blue and black curves, respectively) the probability is close to $0$. In the case of the complete many-orbital Hamiltonian \eqref{HamHubbard} the probability is clearly nonzero, i.e. it is not so rare to find bosons in opposite wells of the potential. \label{Fig4}}
\end{figure}

The time evolution of the probability ${\cal P}(t)$ may shed some light on the differences between evolutions governed by different Hamiltonians in the strong interaction limit. This comes from the fact that for the simplest case \eqref{HamRed}, single-particle tunnelings is strongly suppressed by the conservation of energy. The energy difference between the initial state \eqref{Ini} and a state in which bosons occupy different wells is equal to $U$, and is much larger than the energy gain from the tunneling, $J$. Therefore, the dynamics is governed effectively by the second-order process in the tunneling, i.e. bosons tunnel between wells mainly in pairs \cite{Folling}. This is visible in the evolution of the probability \eqref{Prob} (black curve in Fig. \ref{Fig4}) -- it is close to $0$ at all moments. Situation may change for extended models \eqref{Ham2Mode} and \eqref{HamHubbard} where other processes are taken into account. For example, in the case of the extended two-mode model \eqref{Ham2Mode} the density-induced tunnelings give additional contributions to single-particle tunneling and they effectively support the breaking of a bosonic pair. On the other hand, a pair-tunneling term supports an ordinary second-order tunneling of a composed pair. After all, as it is seen from numerical results, for strong enough interactions the on-site energy $U$ always dominates over other terms and the probability \eqref{Prob} is close to $0$ also for the extended two-mode model \eqref{Ham2Mode} (blue line in Fig. \ref{Fig4}).

The full many-orbital model \eqref{HamHubbard} opens additional channels for single-particle tunneling. For strong interactions, couplings to higher orbitals, in which single-particle tunnelings $J_i$ are large, become relevant. Therefore, the breaking of a bosonic pair is amplified. In consequence, particles can tunnel through the barrier almost independently. This fact is clearly visible in the evolution of the probability ${\cal P}(t)$ (red line in Fig. \ref{Fig4}). The probability, initially being equal to $0$, rapidly grows to $1/2$ and oscillates around $1/4$ during the entire evolution. Consequently, the probability that two bosons may be found in different wells is quite large and cannot be neglected. Moreover, as was explained above, this fact is not reproduced correctly by the simplified models. It is quite natural that for a higher number of particles the situation can be even more complicated. Therefore, one should treat all dynamical results obtained with simplified models with increased care. 

\section{Conclusions}
In this article, we have studied the dynamical properties of two ultra-cold bosons confined in a one-dimensional double-well potential initially occupying the lowest state of a chosen site. We compare the exact dynamics governed by a full two-body Hamiltonian with two simplified two-mode models. In particular, we compared the evolution of particle density and spatial correlations between particles. We show that for a shallow barrier and strong enough interactions the simplified models break down and the correct multi-orbital description cannot be substituted with a two-mode model even if all appropriate interaction terms are taken into account. The fundamental difference between the exact and two-mode descriptions emerges when inter-particle correlations are considered. For example, the evolution of the probability that both bosons are found in opposite wells of the potential crucially depends on couplings to higher orbitals of an external potential. This fact sheds some light on recent theoretical results and opens some perspectives for further experimental explorations.

\section{Acknowledgements} 
The authors thank P. Deuar for a number of very profitable remarks. This work was supported by the (Polish) Ministry of Sciences and Higher Education, Iuventus Plus 2015-2017 Grant No. 0440/IP3/2015/73.

\end{document}